\documentclass{iopart}

\usepackage{iopams}
\usepackage{subfigure}
\usepackage{graphicx
}
\newcommand{\beq}{\begin{equation}}
\newcommand{\eeq}{\end{equation}}
\newcommand{\beqa}{\begin{eqnarray}}
\newcommand{\eeqa}{\end{eqnarray}}
\newcommand{\beqn}{\begin{equation*}}
\newcommand{\eeqn}{\end{equation*}}
\newcommand{\beqan}{\begin{eqnarray*}}
\newcommand{\eeqan}{\end{eqnarray*}}

\newcommand{\ar}{\begin{array}}
\newcommand{\ear}{\end{array}}
\newcommand{\bc}{\begin{color}}
\newcommand{\ec}{\end{color}}
\newcommand{\bit}{\begin{itemize}}
\newcommand{\eit}{\end{itemize}}

\newcommand{\ie}{\emph{i.e.}\ }
\newcommand{\la}{\left\langle}
\newcommand{\ra}{\right\rangle}
\newcommand{\EA}{^e\!\!A}

\newcommand{\text}{\textnormal}

\begin{document}

\title{Isotope shift on the chlorine electron affinity  revisited by an MCHF/CI approach}

\author{T Carette$^{1,2}$ and M R Godefroid$^2$}
\address{$^1$ Department of Physics, Stockholm University, AlbaNova University Centre, SE-106 91 Stockholm, Sweden}
\address{$^2$ Chimie quantique et photophysique, CP160/09, Universit\'e Libre de Bruxelles, B~1050 Brussels, Belgium}
\eads{mrgodef@ulb.ac.be}

\begin{abstract}
Today, the electron affinity is experimentally well known for most of the elements and is a useful guideline for developing \emph{ab initio} computational methods. However, the measurements of isotope shifts on the electron affinity are limited by both resolution and sensitivity. In this context, theory  eventually contributes to the knowledge and understanding of atomic structures, even though correlation plays a dominant role in negative ions properties and, particularly, in the calculation of the specific mass shift contribution. The present study solves the longstanding discrepancy between calculated and measured specific mass shifts on the electron affinity of chlorine~[{\it Phys. Rev. A} {\bf 51}, 231 (1995)].
\end{abstract}

\submitto{\JPB, \today}
\noindent{\it Keywords\/}:
Negative ions, electron affinity, isotope shifts, electron correlation.

\maketitle

\section{Introduction}

Despite the difficulty to measure electron-affinities ($\EA$), experimental values can be considered as ``exact'' by theoreticians for most of the elements. Due to their remarkable properties as weakly bound atomic systems largely governed by electron correlation, negative ions have always attracted broad attention from chemists and atomic physicists~\cite{And:2004a}. With this respect, the field of electron affinities constitutes a playground of choice for testing the computational methods~\cite{deOetal:99a,Kloetal:2010a}. The estimation of the mass polarization operator expectation value that partly makes the level isotope shift~(IS), also highly sensitive to correlation~\cite{Kraetal:87a}, offers other stringent tests for assessing the reliability of theoretical methods. Unfortunately, the mass shifts of transition frequencies or electron detachment thresholds are rather difficult to extract from experimental studies. 
Limitations arise from the size of isotope shifts and/or from the low abundances of the isotopes to be studied, in which cases isotopically enriched samples~\cite{Krietal:09a} or in-source spectroscopy~\cite{CheFla:2010a} could be used. 
One way of deducing a pure mass shift value from observation is the use of King plots~\cite{Kin:84a,Cheetal:2012a}, requiring measurements of isotope shifts on a transition for at least three pairs of isotopes. The nuclear stability valley narrowing for low atomic numbers, there are less possibilities of building the required experimental King plots when approaching the domain of systems for which it is realistic to do highly correlated calculations. Any other way to estimate the field and mass shift contributions separately must contain theoretical inputs~\cite{BauCha:76a,FriHei:2004a}. In cases where either the field shift  (FS) or the mass shift (MS) is a small contribution, the error on the theoretical part is often neglected, although it involves in some cases an unknown systematical error in the data analysis~\cite{Ganetal:2004a}. 

In the context of the physics of highly-charged ions, the balance between the mass and field shift contributions to the level and transition isotope shifts is worthwhile to investigate systematically along an isoelectronic sequence~\cite{Lietal:2012a} for which the increasing importance of a proper relativistic treatment towards the high-$Z$ region becomes obvious~\cite{Sha:85a,Gaietal:2011,Lietal:2012b}. However, for the study of the isotope shifts on the electronic structure of negative and neutral atoms of the second and third periods of the periodic table, electron correlation still largely dominates the relativistic corrections and non-relativistic approaches constitute a good starting point to investigate what makes an isotope shift ``normal'' or ``anomalous''~\cite{GodFro:99b,Nazetal:2012a}.

In the field of negative ions, the laser photodetachment threshold technique  was developed and successfully applied to the measurement of electron affinities~\cite{Neuetal:85a} 
and of the detachment thresholds corresponding to different fine-structure levels of the negative ion and the neutral atom~\cite{Bloetal:06a,Andetal:2007a}. In this method,  the onset of the photodetachment process is measured directly using a tunable laser source. 
In combination with improved methods for the measurement of the photoelectron kinetic energies such as photodetachment microscopy~\cite{Bloetal:96a} and slow-electron velocity-map imaging~\cite{Ostetal:2004a}, the technique of laser photodetachment at threshold represents the most precise way of determining photodetachment thresholds experimentally~\cite{HHS:hrs071}. Amongst the recent applications of photodetachment microscopy, let us cite the first  experiment realized in phosphorus~\cite{Peletal:2011a} with the excitation of the parent neutral atom out of the fundamental spectral term, and the measurement of the electron affinity of selenium with an accuracy of 1~$\mu$eV~\cite{Vanetal:2012a} .
ÜThe possibility of applying the tunable laser photodetachment spectroscopy to the measurement of the isotope shift on the electron affinity has been demonstrated for the first time by Berzinsh~\etal \cite{Beretal:95a} for the $^{35,37}$Cl isotopes. The photodetachment microscopy technique has  been applied later to measure the electron affinities of $^{16}$O and $^{18}$O  separately from a natural sample~\cite{Valetal:99a} and deduce the IS on the oxygen electron affinities for $^{16,18}$O. This work has been extended to $^{17}$O~\cite{Bloetal:01a} with the assistance of theoretical calculations for the estimation of the hyperfine structures. The sulfur electron affinities were measured more recently by photodetachment microscopy for the two isotopes $^{32}$S and $^{34}$S~\cite{Caretal:2010a}, demonstrating the ability of {\it ab initio} methods for estimating the isotope shift on the electron affinity. \\

 More than fifteen years ago, measurements and  many-body calculations were reported for the isotope shift in the chlorine electron affinity in the pioneer work of Berzinsh~\etal \cite{Beretal:95a}. The theory-experiment agreement found for the specific mass shift (SMS) was satisfactory as far as the order of magnitude is concerned but a serious discrepancy lay in its sign. Still, the obtained resolution by tunable-laser photodetachment spectroscopy was remarkable. More recently, the isotope shift on the electron affinity of sulfur was investigated, both experimentally and theoretically~\cite{Caretal:2010a}. It was established that large scale  closed-core multiconfiguration Hartree-Fock (MCHF) calculations lead to a reasonable theory-experiment agreement but it was also shown that core effects are large and cannot be neglected.  The success found in~\cite{Caretal:2010a} for the S/S$^-$ systems incited the present authors to  revisit theoretically the IS on the electron affinity ($^e\!A$) of chlorine, involving only one more electron in the neutral/negative ion balance.  \Sref{sec:th} presents the needed theoretical background, introducing the mass and field isotope shift on the electron affinity.  The correlation models and construction of the configuration spaces are described in \sref{sec:Comp_strat}. \Sref{sec:valres} and \sref{sec:cvres} present, respectively, the results of the valence correlation models and of the open-core configuration interaction (CI) approach. The final theoretical results are summarized and compared with Berzinsh \etal's experimental and many-body calculations results~\cite{Beretal:95a} in~\sref{sec:comp}.

\section{Electron-affinity and isotope shifts}\label{sec:th}

\subsection{The experimental electron affinity as a guideline}
\label{sec:EA_guideline}

Like in our previous work on the isotope shift in the sulfur electron affinity~\cite{Caretal:2010a}, the experimental electron affinity 
\begin{equation}
\label{eq:EA}
\EA_{\text{\scriptsize exp}} = E(\text{Cl} \; 3s^2 3p^5 \;  ^2P^o_{3/2}) -  E(\text{Cl}^- \; 3s^2 3p^6 \;  ^1S_{0})
\end{equation}
is used as a precious guideline to set efficiently pathways in the variational configuration spaces.
Both $^{35}$Cl and $^{37}$Cl isotopes have a nuclear spin $I=3/2$. For both isotopes, the ground state of the neutral chlorine is therefore split into four hyperfine structure levels $F=0,1,2,3$ where the $F=0$ level has the lowest energy,  while the ground state of the negative ion does not show hyperfine structure since $J=0$.
  The  electron affinity measured by Berzinh \etal~\cite{Beretal:95a,And:2004a},
\[ \EA_{\text{\scriptsize exp}} = 29~138.59(22)~\mbox{cm}^{-1} = 3.612~724(27)~\mbox{eV} \]
is defined from the threshold energy and thus strictly corresponds to the difference between the lowest hyperfine level of the ground state of the neutral and the ground state of the anion. This represents a difference with eq.~\eref{eq:EA} of only  700 and 586~MHz for $^{35}$Cl and $^{37}$Cl, respectively and is one order of magnitude smaller  than the uncertainty of the measured electron affinity.
In the present work, we adopt a non-relativistic variational approach for targeting electron correlation (see Section~\ref{sec:Comp_strat}) and we need to estimate the average experimental electron affinity that would be measured if not resolving the fine-structure thresholds due to the $J$-splitting of the chlorine ground term Cl~$3s^2 3p^5 \; ^2P^o_{1/2-3/2}$. This average electron affinity can be expressed as
\begin{eqnarray}
\EA_{\text{\scriptsize exp}}^\text{\scriptsize AV} & = &
\overline{E}(\text{Cl} \; ^2P^o ) - \overline{E}(\text{Cl}^- \; ^1S) \nonumber \\
& = & \frac{(4 E_{3/2} + 2 E_{1/2})}{6} - E(\text{Cl}^- \; ^1S_{0})  \\
& = & \EA_{\text{\scriptsize exp}} + \frac{E_{1/2} - E_{3/2}}{3} \; . \nonumber
\end{eqnarray}
Using the NIST fine structure energy separation $(E_{1/2} - E_{3/2} = +882.3515~\mbox{cm}^{-1})$, one finds $ \EA_{\text{\scriptsize exp}}^\text{\scriptsize AV} = 29~432.70(22)~\mbox{cm}^{-1} = 3.649~189~\mbox{eV}$, from which one substracts the non-fine structure contribution $\Delta E^\text{\scriptsize NF}= -0.01509~\mbox{eV}$ reported by de Oliveira {\it et al}~\cite{deOetal:99a} to estimate the reference non-relativistic electron affinity
\begin{equation}
\label{eq:EA_ref}
 \EA_{\text{\scriptsize ref}}^\text{\scriptsize NR} =  \EA_{\text{\scriptsize exp}}^\text{\scriptsize AV}  - \Delta E^\text{\scriptsize NF} = 3.664~28~\mbox{eV}
 = 0.134~660~\mbox{E}_h \; .
\end{equation}

Adopting the ($A' > A$) convention where $A$ is the mass number, the isotope shift (IS) on the $\EA$ is defined as
\beq
\label{IS_EA}
IS(A',A) = \delta \EA \equiv  \EA(A') -  \EA(A)\, .
\eeq
It can be expressed as the sum of the mass shift (MS) and field shift (FS) contributions
\beq
 \delta \EA =  \delta \EA_{\text{\scriptsize MS}} + \delta \EA_{\text{\scriptsize FS}} .
\eeq

\subsection{Isotope Mass Shift}
\label{sec:MS}

The  mass shift of an atomic energy level is the energy displacement due to the inclusion of the dynamics of the nucleus in the Hamiltonian.
The proper non-relativistic quantum mechanics treatment of the separation of the center of mass motion and the motion of the $N$ electrons relative to the nucleus can be found in Johnson~\cite{Joh:2007a} and  Bransden and Joachain~\cite{BraJoa:83a}.
A first (one-body) effect of the finite nuclear mass is to scale the infinite mass energies by the ratio
$\mu/m_e = M_A/(M_A + m_e)$. The shift of the energy from the infinite-mass value defines the normal mass shift (NMS)
\beqa
\label{LNMS}
\delta E_{\text{\sc{nms}}} = E_M - E_\infty = - \frac{m_e}{M_A + m_e} E_\infty =
- \frac{\mu}{M_A} E_\infty =  - \frac{m_e}{M_A} E_M \; .
\eeqa
A second (two-body) correction to the energy, referred to as the specific mass shift (SMS) is given by
\beqa
\label{LSMS}
\hspace*{-1.5cm} \delta E_{\text{\sc{sms}}} = \frac{M_A}{2(M_A + m_e)^2} 
\left\langle \sum_{i \neq j}^N {\bf p}_i \cdot {\bf p}_j \right\rangle 
=\frac{\mu_A}{(M_A + m_e)} \frac{1}{m_e} 
\left\langle \sum_{i < j}^N {\bf p}_i \cdot {\bf p}_j \right\rangle \; .
\eeqa
Combining \eref{LNMS} and \eref{LSMS} for estimating the level mass isotopic  shift ($\delta E_{\text{\sc{ms}}} = \delta E_{\text{\sc{nms}}} + \delta E_{\text{\sc{sms}}}$), it is easy to obtain the expression of the mass contribution to the isotope shift on the electron affinity~\eref{IS_EA}
\beq
\label{NMS_plus_SMS}
 \delta \EA_{\text{\scriptsize MS}}  = 
 \delta \EA_{\text{\scriptsize NMS}}   +   \delta \EA_{\text{\scriptsize SMS}}   
\eeq
\[
=  \left( - \frac{\mu'}{M'} + \frac{\mu}{M}\right) \ ^e\!A_\infty 
+ \left( \frac{\mu'}{M'+ m_e} -  \frac{\mu}{M+ m_e} \right)\ \left( \frac{\hbar^2}{m_e} \right) \Delta S_{\text{\sc{sms}}} 
\]
where $\Delta S_{\text{\sc{sms}}} $ is the difference 
\begin{equation}
\label{Delta_SMS}
\Delta S_{\text{\sc{sms}}} =   S_{\text{\sc{sms}}} (X) -  S_{\text{\sc{sms}}} (X^-) \;
\end{equation}
of the expectation values defining the SMS parameters
\begin{equation}
\label{eq:Ssms}
S_{\text{\sc{sms}}} = 
- \left\langle \Psi_\infty  \left| \sum_{i<j}^N \nabla_i \cdot \nabla_j \right| \Psi_\infty  \right\rangle \; , 
\end{equation} 
calculated for the ground states of the neutral atom $X$ and the negative ion $X^-$.
Note that \eref{NMS_plus_SMS} can be shown to be strictly equivalent to
\beqa   \delta \EA_{\text{\scriptsize MS}}  
   &=& \left( \frac{\mu}{M} - \frac{\mu'}{M'}\right) \left( \EA_\infty - \frac{\hbar^2}{m_e}\Delta S_{\text{\sc{sms}}} \right)\nonumber \\&&
+ \left[ \Big(\frac{\mu}{M}\Big)^2-\Big(\frac{\mu'}{M'}\Big)^2\right]  \frac{\hbar^2}{m_e}\Delta S_{\text{\sc{sms}}}\, , \label{eq:ea_sep}
\eeqa
where the first term alone corresponds to the prescription of King~\cite{Kin:84a}\footnote{The second term arises from the all-order treatment of the NMS (see also footnote~(5) of \cite{Godetal:01a}).}.

The expectation value~\eref{eq:Ssms} is dominated by its Hartree-Fock value \emph{i.e.}, in Cl, by the $(s,p)$ Vinti integrals~\cite{Vin:39a} which contribute negatively. The differential SMS~\eref{Delta_SMS}  between the neutral atom and the negative ion is itself dominated by the Vinti integrals involving the  additional $3p$ electron in Cl$^-$. In general, the mass polarization difference $\Delta S_{\text{\sc{sms}}}$ is therefore positive, as expected. The first term  appearing in the mass shift on the electron affinity~\eref{eq:ea_sep} explicits the antagonism between the two positive contributions, ie. the electron affinity $\EA_\infty$ by itself and the SMS parameter variation $(\hbar^2/m_e) \Delta S_{\text{\sc{sms}}}$, from which strong cancellation between the electronic part of the NMS and SMS can be expected~: classically, electrons tend to move in opposite directions, as beautifully illustrated by Krause \etal~\cite{Kraetal:87a} in the two-electron case  and the nucleus to be slower that expected from the one-electron contributions.
The SMS term is usually the bottleneck in the calculation of the isotope shift of light systems and most of our effort is aimed at a robust estimation of this effect in the non-relativistic level of approximation.

The nuclear masses have been estimated by substracting the total electrons mass from the atomic masses of $^{35}$Cl and $^{37}$Cl taken from~†\cite{Audetal:03a}, ie. $M(^{35}\mbox{Cl}) =  34.95952682(4)$~u and $M(^{37}\mbox{Cl}) = 36.95657673(5)$~u.

\subsection{Isotope Field Shift}
\label{sec:MS}\label{sec5:FS}

According to~\cite{HeiSte:74a,Toretal:85a}, 
the field shift on the electron affinity can be  estimated from the following expression
\beq
\label{eq:IS_EA}
\delta \EA_{\textsc{fs}} = 4\pi \left[ \rho({\bf 0})^X_{\textsc{nr}} - \rho({\bf 0})^{X^-}_{\textsc{nr}} \right] \frac{ha_0^3}{4Z}  f(Z)^{AA'} \left[ \left\langle r^2\right\rangle_{A'} - \left\langle r^2\right\rangle_{A} \right]
\eeq
where $\rho_{\textsc{nr}}({\bf 0})$ is the spin-less total electron density~\cite{Boretal:2010a} at the origin, and where $f(Z)$ corrects for the fact that we use the non-relativistic electronic density for a point nucleus. Its value  $f(Z=17)^{35, 37}=16.940~\mbox{mK/fm}^2$ $= 507.8~\mbox{MHz/fm}^2$ 
is taken from Aufmuth \etal's compilation~\cite{Aufetal:87a}.

The knowledge on the nuclear charge distributions of chlorine isotopes is unfortunately limited~\cite{FriHei:2004a}. The root mean square nuclear charge radii $ \langle r^2 \rangle^{1/2} $ resulting from elastic electron scattering measurements have been reported by Briscoe \etal~\cite{Brietal:80a} for both isotopes
\begin{eqnarray*}
\label{Briscoe_rms}
\langle r^2 \rangle^{1/2}_{37}&=&3.384(15)~\text{fm} \; ,\\
\langle r^2 \rangle^{1/2}_{35}&=&3.388(15)~\text{fm} \; ,
\end{eqnarray*}
on the basis of phase-shift fits to a three-parameter Fermi distribution. For  $^{35}$Cl, the latter value strongly disagrees with the $\langle r^2 \rangle^{1/2}_{35}=3.335(18)$~fm rms-value extracted from earlier muonic X-ray measurements~\cite{Bacetal:67a}.
To evaluate the difference $\delta \la r^2\ra^{35,37} = \la r^2\ra_{37} - \la r^2\ra_{35}$ appearing in \eref{eq:IS_EA}, we choose for consistency, the two above values recommended by Briscoe \etal~\cite{Brietal:80a} and also adopted in De Vries {\it et al}'s compilation~\cite{DeVetal:87a}, ie. 
\beq
\delta \la r^2\ra^{35,37} = -0.03(24)~\text{fm}^2 \; .
\eeq
Note that Berzinh \etal  used for their FS estimation~\cite{Beretal:95a} the value of 
$\delta \la r^2\ra^{35,37} = +0.12(12)$~fm$^2$  that can also be derived  from Briscoe \etal's charge distribution parameters, but obtained  using  two-parameter Fermi models. This positive value is in line with the result 
$\delta \la r^2\ra^{35,37} = +0.127$~fm$^2$ that can be calculated from the rms of Angeli's compilation~\cite{Ang:2004a} that have been determined by averaging the charge radii extracted from different sources~\cite{Ang:99a}. In view of the large dispersion of the nuclear data, we think that it is preferable to use the same reference for both isotopes to maintain some coherence. 

\section{Computational strategy}
\label{sec:Comp_strat}

\subsection{MCHF calculations}
\label{sec:MCHF}

We use multiconfigurational Hartree-Fock (MCHF) wave functions built on orthonormal numerical radial orbitals using the \textsc{atsp2k} package \cite{Froetal:2007a}. Such an expansion is written
\beq
\Psi^{\textsc{mchf}}(\Gamma LS) = \sum_i  c_i \;  \Phi_i(\gamma_i LS) \label{eq:mchfwfn} \; ,
\eeq
where the configuration state functions (CSF) $\Phi_i (\gamma_i LS)$ are symmetry-adapted linear combinations of Slater determinants. In MCHF theory, both one-electron radial functions $\{ P_{nl}(r) \}$ intervening in the construction of the CSF's and the interaction coefficients $\{ c_i \}$ are optimized~\cite{Fro:77a,Fro:86a}, while the orbitals are fixed in configuration interaction (CI) calculations~\cite{Froetal:97b}.

\subsection{Construction of the CSFs spaces}
\label{sec:csf}

First, a multireference configuration set (MR) is selected. Then, the ``multireference-interacting CSF space'' (MR-I) is generated as the set of all CSF's interacting with at least one CSF of the MR, ie.
\beq\label{eq:MR-I}
\Phi_i(\gamma_i\ LS) \in \text{ MR-I }
\Leftrightarrow
\exists \Phi_k \in \text{ MR with }  \left\langle\gamma_i\ LS \left| H \right| \gamma_k\ LS \right\rangle \neq 0 
\eeq
whatever the one-electron radial functions are.
For a given $n_{max}l_{max}$ defining the orbital active space $\lceil n_{max}l_{max} \rceil$ of subshells with $n\leq n_{max}$ and $l\leq l_{max}$, the MR-I$\lceil n_{max}l_{max} \rceil$ set  is a subset of the single and double excitations of the MR (MR-SD$\lceil n_{max}l_{max} \rceil$). 
As advocated in Ref.~\cite{Caretal:2010a}, we choose to couple the subshells in decreasing order of $n$ and $l$.

\subsection{Valence CSF spaces}\label{sec:valcsf}

Keeping the [Ne] core ($1s^2 2s^2 2p^6$) closed, we use the following multireference
\beq
\text{MRD} = \text{[Ne]}\{3s, 3p\}^6 \{3, 4\}^2 \label{eq:MRCl-}
\label{MRD}
\eeq
for Cl$^-$, allowing two electrons to ``leave'' the spectroscopic subshells $3s$ and $3p$ and to be excited in the orbitals of the $n=3, 4$ shells. Since this list is limited to double excitations of the main configuration, the  MRD-I expansions contain at most quadruple excitations. Heeding the strong $3p^2 \rightarrow 3d^2, 4p^2$ interactions, we also explore the MR constructed by merging the MRD~\eref{MRD} with the single excitations of the two configurations $3s^23p^43d^2$ and $3s^23p^44p^2$ in the shells $n=3$, $4$. This list, denoted MRT, allows quintuple excitations in the final MRT-I lists.

For Cl, we adopt two different approaches, leading to significantly different $\{P_{nl}(r)\}$ sets. Firstly, we select only the most important configurations in the multireference, omitting the single excitations conserving~$l$
\begin{eqnarray}
\text{MR}1_4&=&\text{[Ne]}\{ 3s^2 3p^5, 3s^2 3p^3 3d^2, 3s^2 3p^3 4p^2, 3s^1 3p^5 3d^1\}\, ,\\
\text{MR}1_5&=& \text{MR}1_4 \cup \{ 3s^1 3p^4 3d^1 4f^1\}\, .
\end{eqnarray}
Secondly, we use a systematic approach based on the following MR space
\beq
\text{MR2} =\text{[Ne]} \{3s, 3p\}^5 \{3\}^1 \{3, 4\}^1\, .\label{eq:MRCl1}
\eeq
MCHF calculations are performed for various $n_{max}l_{max}$, ranging from $4f$ to $11k$, the core orbitals being either optimized (MCHF) or fixed to their HF shape (FC-MCHF). Table~\ref{tab:w} presents the configuration weights obtained in the FC-MCHF calculations. These weigths are defined as the summed contribution of all CSFs $\Phi_i$  belonging to a given configuration:
\begin{equation}
\label{weights}
w = \sqrt{\sum_{\Phi_i \in \{config\}} c_i^2} \; .
\end{equation}
\Table{Weights of the configurations composing the MR1$_4$, MR1$_5$ and MR2 sets of Cl and the MRD and MRT sets of Cl$^-$ in the corresponding MR-I$\lceil 11k\rceil$ wave functions with the core-orbitals fixed at the HF approximation. \# is the index of each configuration.\label{tab:w}}
\footnotesize
\begin{tabular}{cccc|r}
\br
 \multicolumn{5}{c}{Cl $3p^5$ $^2P^o$}\\
\mr
 \multicolumn{2}{c}{MR1$_4$}  &  \multicolumn{2}{c|}{MR2} \\
config. & $w$ & config. & $w$ & \#  \\
 \mr
$3s^2 3p^5          $& 0.9569 &%
$3s^2 3p^5          $& 0.9374 & 1 \\
$3s^23p^33d^2       $& 0.1903 &%
$3s^23p^33d^2       $& 0.1835 & 2 \\
$3s^13p^53d^1       $& 0.1143 &%
$3s^23p^44p^1       $& 0.1792 & 3 \\
$3s^23p^34p^2       $& 0.0831 &%
$3s^13p^53d^1       $& 0.1098 & 4 \\
&&%
$3s^13p^54s^1       $& 0.0910 & 5 \\
&&%
$3s^13p^43d^14f^1   $& 0.0690 & 6 \\
\cline{1-2}
\multicolumn{2}{c}{MR1$_5$}&%
$3s^13p^43d^14p^1   $& 0.0629 & 7 \\
config. & $w$&%
$3s^23p^44f^1       $& 0.0485 & 8 \\
\cline{1-2}
$3s^2 3p^5            $& 0.9562 &%
$3p^53d^2             $& 0.0427 & 9 \\
$3s^23p^33d^2       $& 0.1900 &%
$3s^23p^33d^14d^1   $& 0.0382 & 10\\
$3s^13p^53d^1       $& 0.1140 &%
$3p^64p^1             $& 0.0170 & 11\\
$3s^23p^34p^2       $& 0.0831 &%
$3s^13p^54d^1       $& 0.0150 & 12\\
$3s^13p^43d^14f^1   $& 0.0704 &%
$3p^53d^14s^1       $& 0.0074 & 13\\
&&%
$3p^53d^14d^1       $& 0.0062 & 14\\
&&%
$3s^23p^33d^14s^1   $& 0.0014 & 15\\
\br
 \multicolumn{5}{c}{Cl$^-$ $3p^6$ $^1S$}\\
\mr
 \multicolumn{2}{c}{MRD}  &  \multicolumn{2}{c|}{MRT} \\
config. & $w$ & config. & $w$ & \# \\
 \mr
$3s^2 3p^6          $& 0.9469 &%
$3s^2 3p^6          $& 0.9124 & 1\\
$3s^23p^43d^2       $& 0.2090 &%
$3s^23p^54p^1       $& 0.2592 & 2\\
$3s^23p^44p^2       $& 0.1283 &%
$3s^23p^43d^2       $& 0.2066 & 3\\
$3s^13p^54s^14p^1   $& 0.0737 &%
$3s^23p^44p^2       $& 0.0876 & 4\\
$3s^13p^53d^14f^1   $& 0.0711 &%
$3s^13p^64s^1       $& 0.0849 & 5\\
$3s^13p^53d^14p^1   $& 0.0639 &%
$3s^13p^53d^14p^1   $& 0.0769 & 6\\
$3s^23p^43d^14d^1   $& 0.0528 &%
$3s^13p^53d^14f^1   $& 0.0738 & 7\\
$3s^13p^54p^14d^1   $& 0.0517 &%
$3s^13p^54s^14p^1   $& 0.0539 & 8\\
$3s^23p^44f^2       $& 0.0455 &%
$3s^23p^33d^24p^1   $& 0.0535 & 9\\
$3s^23p^44d^2       $& 0.0399 &%
$3s^23p^44f^2       $& 0.0448 &10\\
$3s^23p^44s^14d^1   $& 0.0364 &%
$3s^23p^43d^14d^1   $& 0.0441 &11\\
$3s^23p^54p^1       $& 0.0355 &%
$3s^23p^44s^14d^1   $& 0.0413 &12\\
$3s^23p^44s^2       $& 0.0292 &%
$3s^23p^44d^2       $& 0.0375 &13\\
$3p^63d^2           $& 0.0272 &
$3p^63d^2           $& 0.0348 &14\\
$3s^13p^54d^14f^1   $& 0.0261 &%
$3s^23p^34p^3       $& 0.0298 &15\\
$3p^63d^14d^1       $& 0.0232 &%
$3s^23p^44s^2       $& 0.0291 &16\\
$3p^64p^2           $& 0.0207 &%
$3s^13p^54p^14d^1   $& 0.0256 &17\\
$3s^23p^43d^14s^1   $& 0.0203 &%
$3s^23p^33d^24f^1   $& 0.0252 &18\\
$3s^23p^44p^14f^1   $& 0.0190 &%
$3s^13p^44s^14p^2   $& 0.0234 &19\\
$3p^64s^2           $& 0.0178 &%
$3s^13p^43d^24s^1   $& 0.0232 &20\\
$3p^64d^2           $& 0.0140 &%
$3p^64p^2           $& 0.0209 &21\\
$3p^64f^2           $& 0.0126 &%
$3s^23p^44p^14f^1   $& 0.0182 &22\\
$3s^13p^64s^1       $& 0.0046 &%
$3p^64s^2           $& 0.0138 &23\\
&&%
$3s^13p^43d^14p^2   $& 0.0137 &24\\
&&%
$3p^64f^2           $& 0.0122 &25\\
&&%
$3s^13p^43d^3       $& 0.0115 &26\\
&&%
$3p^63d^14d^1       $& 0.0107 &27\\
&&%
$3s^13p^4	3d^24d^1   $& 0.0087 &28\\
&&%
$3s^13p^4	4p^24d^1   $& 0.0070 &29\\
&&%
$3s^23p^43d^14s^1   $& 0.0069 &30\\
&&%
$3p^64d^2           $& 0.0044 &31\\
&&%
$3s^13p^54d^14f^1   $& 0.0026 &32\\
&&%
$3s^23p^34p^24f^1   $& 0.0021 &33\\
\br
\end{tabular}
\endTable
\Table{Values of $p$ chosen to perform the corresponding open-core CI calculations for each MR of Cl and Cl$^-$. \label{tab:ps}}
\begin{tabular}{lcc}
\br
&MR & selected $p$\\
\mr
&MRD     & 1$-$6, 8, 10, 13, 14, 16, 19, 23\\
\raisebox{1ex}[0cm]{Cl$^-$} &MRT     & 1$-$9, 12, 14, 16, 20, 22, 27, 33\\
\mr
&MR1$_4$ & 1$-$4\\
Cl&MR1$_5$ & 1$-$5\\
&MR2     & 1$-$10, 12, 15\\
\br
\end{tabular}
\endTable
Contrary to the MR$1_4$(Cl), MR$1_5$(Cl) and MRD(Cl$^-$) lists, the use of MR2(Cl) and MRT(Cl$^-$) reference lists favors the apparition of large weights for the single substitutions $3s\rightarrow 4s$ and $3p\rightarrow 4p$ by a mechanism identified by the authors~\cite{Carette_PhD:2010a}.
The importance of these configurations comes from the strong interaction of the total wave function with the Brillouin state of the main configuration, permitted by a large deviation of the orbitals from their HF shapes. Very similar wave functions can be constructed with and without predominance of the $nl \rightarrow n'l$ single excitations, the limit case occurring when the wave function is invariant under the rotations between the orbitals of a given $l$. When these single excitations dominate the expansion, the odd $3l \rightarrow 4l$ ($l=s,p$) excitations -- \ie the substitution of an odd number of $3l$ electrons by $4l$ electrons -- are favored while the even ones are suppressed. In particular, this means that the MR2-I model can mimic, thanks to the orbital optimization, the presence of the $3p^2 \rightarrow 4p^2$ excitation in the reference. This comes with the price of relatively ``dilute'' eigenvectors and a slower numerical convergence of the MCHF calculations. Other manifestations of this effect can be found in Ref.~\cite{Caretal:2010a}.

\subsection{Core-Valence CSF spaces}\label{sec:cvcsf}

Following the hierarchy of Table~\ref{tab:w}, subsets of $p$ configurations, denoted MR$_p$, are selected. For each $p$, all the single and double subshell substitutions of the corresponding MR$_p$ in $\lceil 11k\rceil$ are generated, allowing at most one hole in the $n=2$ shell and none in $1s$.
These lists are finally reduced with respect to the full MR according
 according to the building rule~\eref{eq:MR-I}.
The resulting space is denoted MR-I/CV$\! p$.
The $p$ values selected for each MR~(\ref{eq:MRCl-}-\ref{eq:MRCl1}) are presented in Table~\ref{tab:ps}.

\section{Results and discussion}

\subsection{Valence results}\label{sec:valres}

The results of the MCHF and FC-MCHF calculations using the multireferences MRD for Cl$^-$, and MR2 for Cl, with $n_{max}l_{max}=9k, 10k$ and $11k$ are presented in Table~\ref{tab:val1}. The $\EA$ and $S_{\text{\sc{sms}}}$ values converge smoothly with the number of correlation layers but a significant difference is found between the MCHF and FC-MCHF approaches. The ground state of Cl$^-$ is more stabilized by the optimization of the core orbitals than the one of Cl. The impact on the difference of mass polarization parameters is enormous (\mbox{$\approx 36\%$}), leading to a variation of the isotope shift of about~\mbox{$70\%$}.
The MCHF results of Table~\ref{tab:val1} are in agreement with the value of the differential mass polarization \mbox{$\Delta S_{\text{\sc{sms}}} = 0.091(25)$ $a_0^{-2}$} extracted from the experiment~\cite{Beretal:95a} and the $\EA_{\text{\scriptsize ref}}^\text{\scriptsize NR}$ value~\eref{eq:EA_ref}.

\fulltable{Total energies ($E$, in E$_h$), $S_{\text{\sc{sms}}}$ parameters (in $a_0^{-2}$) and size of the expansions (NCSF) in the MRD-I model of Cl$^-$ and in the MR2-I model of Cl for the largest active sets. The core-orbitals are either fixed to their HF shapes (FC-MCHF) or optimized (MCHF). The electron affinity ($\EA$) and $\Delta S_{\text{\sc{sms}}}$ calculated using these models are reported in the third part of the table. \label{tab:val1}}
\footnotesize
\begin{tabular}{lccrrrr}
\br
& & & \multicolumn{2}{c}{FC-MCHF} & \multicolumn{2}{c}{MCHF} \\
&&&\crule{2}&\crule{2}\\
& $nl$ max & \multicolumn{1}{c}{NCSF} & \multicolumn{1}{c}{$E$}  & \multicolumn{1}{c}{$S_{\text{\sc{sms}}}$} & \multicolumn{1}{c}{$E$}  & \multicolumn{1}{c}{$S_{\text{\sc{sms}}}$} \\
\mr
            & $ 9k$ & 113~691 & -459.838~5962 & -80.445~6256 & -459.839~4636 & -80.420~4925\\
Cl$^-$(MRD) & $10k$ & 159~948 & -459.838~8245 & -80.445~8813 & -459.839~6907 & -80.420~8569 \\
            & $11k$ & 214~416 & -459.838~9311 & -80.446~0807 & -459.839~7916 & -80.421~0037 \\
\mr
       & $ 9k$ & 291~878 & -459.704~5818 & -80.324~8911 & -459.704~9422 & -80.331~9936\\
Cl(MR2)& $10k$ & 410~462 & -459.704~7414 & -80.325~1537 & -459.705~0994 & -80.332~2030\\
       & $11k$ & 550~117 & -459.704~8192 & -80.325~3422 & -459.705~1767 & -80.332~3553\\
\mr
& \multicolumn{2}{c}{$nl$ max} & \multicolumn{1}{c}{$\EA$}   & \multicolumn{1}{c}{$\Delta S_{\text{\sc{sms}}}$} & \multicolumn{1}{c}{$\EA$} & \multicolumn{1}{c}{$\Delta S_{\text{\sc{sms}}}$} \\
\mr
    &\multicolumn{2}{c}{$9k$}       & 0.134~0144 & 0.120~7345 & 0.134~5215 & 0.088~4989 \\
Diff&\multicolumn{2}{c}{$10k$}      & 0.134~0831 & 0.120~7276 & 0.134~5913 & 0.088~6538\\
    &\multicolumn{2}{c}{$11k$}      & 0.134~1119 & 0.120~7384 & 0.134~6149 & 0.088~6483\\
 \mr
Exp. & & & & & 0.134~7(1)\parbox{0.05cm}{$^a$} & 0.091(25)\hspace{0.17cm}~\\

 \br
\multicolumn{7}{l}{\footnotesize $^a$ $\EA_{\text{\scriptsize ref}}^\text{\scriptsize NR}$ (see Eq.~\ref{eq:EA_ref}).}
 \end{tabular}
\endfulltable

It should however be noted that, due to the total wave function symmetry properties, the core optimization in the MCHF calculations leads to significantly different contributions in Cl$^-$ and Cl. Indeed, the equivalence of the $2p$ and $3p$ subshells in the main configuration of Cl$^-$ allows a large rotation of the $P_{2p}(r)$ and $P_{3p}(r)$ radial functions as it leaves the dominant contribution to the wave function ($1s^22s^22p^63s^23p^6$) invariant. Contrary to the anion, this effect is suppressed  in the neutral atom by the smallness of the $2p^63p^5 \leftrightarrow 2p^53p^6$ interaction.

\begin{figure}
\center
\caption{
Top: total radial density of the Cl$^-$ MCHF wave function calculated with the program {\sc density}~\cite{Boretal:2010a} and its [Ne] core and valence contributions. It is compared to the valence radial density of neutral chlorine.
Bottom: total difference of radial density between Cl$^-$ and Cl and its core and valence contributions.\label{fig:dens}}
\includegraphics[width=\textwidth]{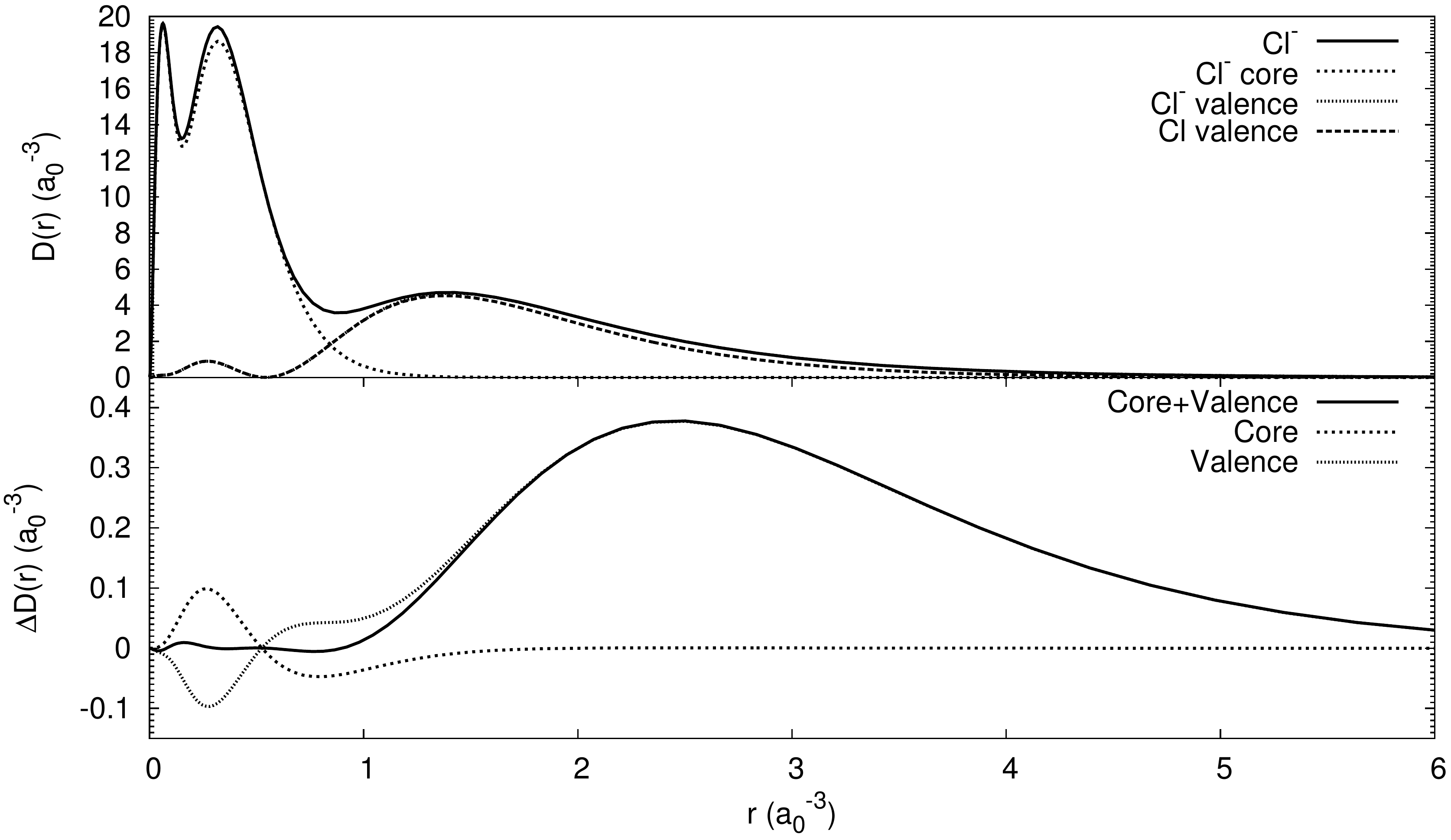}
\end{figure}

Figure~\ref{fig:dens} illustrates the dramatic mixing of core and valence occurring in Cl$^-$ as compared to Cl. The plot at the top shows the total radial density  calculated with the program {\sc density}~\cite{Boretal:2010a} from the Cl$^-$ MCHF wave function. In a closed core calculation, this total density can easily be separated into its [Ne] core contribution and its valence contribution. It is compared to the valence radial density of neutral chlorine.  In the bottom panel of Figure~\ref{fig:dens}, we present the total difference of radial density between Cl$^-$ and Cl as well as its core and valence contributions. It is usually accepted that the core could change only slightly when comparing the neutral atom with its negative ion~\cite{And:2004a}. This is indeed observed in the total radial density difference, the ``extra" electron of Cl$^-$ being efficiently screened out of the [Ne] core. However, we can also observe that it is the result of a large compensation between the core and valence contributions. It means that ``freezing the core" does not mean the same thing in Cl and Cl$^-$.
This effect did not appear when comparing S$^-$ to  neutral sulfur~\cite{Caretal:2010a}. 

The impact of the Cl$^-$ main configuration symmetry on the optimal core orbitals is small enough  that MCHF results can still be considered meaningful although \emph{a priori} untrustworthy from a strict point of view. Nevertheless the lack of a clean distinction between  core and valence contributions to the wave function becomes dramatic in an open-core CI approach. 

Table~\ref{tab:val2} presents the FC-MCHF calculations MR1$_4$-I, MR1$_5$-I of Cl and MRT of Cl$^-$ for $\lceil 9k\rceil$, $\lceil 10k\rceil$ and $\lceil 11k\rceil$. Note that, from the comparison of Tables~\ref{tab:val1}  and~\ref{tab:val2}, the energy values of the MR1$_5$-I and MR2-I models are in close agreement but the corresponding $S_{\text{\sc{sms}}}$ parameters do not compare so well, the mass polarization expectation values of the MR2 based approach being in better agreement with the MR1$_4$-I model. It also should be observed that the MR2-I CSF spaces are almost twice larger than the ones of  MR1$_5$-I for a given number of correlation layers.

\Table{Total energies ($E$, in E$_h$), $S_{\text{\sc{sms}}}$ parameters (in $a_0^{-2}$) and size of the expansions (NCSF) in the MRT-I model for Cl$^-$, in the MR1$_4$-I and MR1$_5$ models for Cl. The active set is limited to $9k$, $10k$ and $11k$ and the core orbitals are fixed to their HF shapes.\label{tab:val2}}
\begin{tabular}{lcrcc}
\br
& & & \multicolumn{2}{c}{FC-MCHF}\\
&&&\crule{2}\\
& $nl$ max & \multicolumn{1}{c}{NCSF} & \multicolumn{1}{c}{$E$}  & \multicolumn{1}{c}{$S_{\text{\sc{sms}}}$}\\
\mr
            & $ 9k$ & 330~083 & -459.839~5099 & -80.432~1726\\
Cl$^-$(MRT) &$ 10k$ & 467~617 & -459.839~7392 & -80.432~4083\\
            &$ 11k$ & 630~131 & -459.839~8463 & -80.432~6017\\
\mr
            & $ 9k$ &  65~286 & -459.703~5158 & -80.324~7147\\
Cl(MR1$_4$) & $10k$ &  91~915 & -459.703~6706 & -80.324~9490\\
            & $11k$ & 123~357 & -459.703~7494 & -80.325~1612\\
\mr
            & $ 9k$ & 152~483 & -459.704~4618 & -80.323~3889\\
Cl(MR1$_5$) & $10k$ & 212~132 & -459.704~6178 & -80.323~6181\\
            & $11k$ & 282~047 & -459.704~6972 & -80.323~8321\\
\br
\end{tabular}
\endTable

\subsection{Open-core CI analysis and error estimation}\label{sec:cvres}

In the following, we denote the Cl and Cl$^-$ models $(r,p)$ and $(r',p')$, respectively, $r$ denoting the used multireference, \emph{i.e.} either $1_4,1_5,2$ for Cl or $D,T$ for Cl$^-$.
The full array of results of the open-core CI calculations on Cl and Cl$^-$ are reported in Table~\ref{tab5:fullCV}. 
\Table{Full array of open-core CI $(r,p)$ results for Cl and Cl$^-$ (see text). All values are given in in atomic units.\label{tab5:fullCV}}
\footnotesize
\begin{tabular}{lrccclrcc}
\br
 \multicolumn{4}{c}{Cl $3p^5 \; ^2P^o$} && \multicolumn{4}{c}{Cl$^-$ $3p^6 \; ^1S$}\\
 \crule{4}&&\crule{4}\\
$r$ & $p$ & \multicolumn{1}{c}{$E$} & \multicolumn{1}{c}{$S_{\text{\sc{sms}}}$}&&$r'$ & $p'$ & \multicolumn{1}{c}{$E$} & \multicolumn{1}{c}{$S_{\text{\sc{sms}}}$}\\
\mr
1$_4$&$1$ & -459.7637550 & -79.841847 && D&$1$& -459.8967475 & -79.984407 \\
&$2$      & -459.7665316 & -79.769342 && &$2 $& -459.8999568 & -79.895440 \\
&$3$      & -459.7672007 & -79.768308 && &$3 $& -459.9013473 & -79.877546 \\
&$4$      & -459.7677966 & -79.760567 && &$5 $& -459.9022500 & -79.871812 \\
&         &              &            && &$6 $& -459.9024769 & -79.870302 \\
1$_5$&$1$ & -459.7646251 & -79.841488 && &$8 $& -459.9028466 & -79.869516 \\
&$2$      & -459.7674644 & -79.763679 && &$10$& -459.9030490 & -79.870592 \\
&$3    $  & -459.7681956 & -79.762271 && &$13$& -459.9031475 & -79.869808 \\
&$4    $  & -459.7687944 & -79.754529 && &$14$& -459.9031978 & -79.869350 \\
&$5     $ & -459.7691127 & -79.754750 && &$16$& -459.9032499 & -79.869245 \\
1$_6$&$6$ & -459.7696487 & -79.752266 && &$19$& -459.9033102 & -79.867757 \\
&         &              &            && &$23$& -459.9033423 & -79.867577 \\
2&$1$     & -459.7638697 & -79.820005 && \\
&$2 $     & -459.7667435 & -79.735797 &&T& $1$& -459.8927326 & -80.042016 \\ 
&$3 $     & -459.7679924 & -79.762100 && &$2 $& -459.8984329 & -79.952901 \\
&$4 $     & -459.7686643 & -79.756779 && &$3 $& -459.9015022 & -79.876465 \\
&$5 $     & -459.7688295 & -79.758733 && &$4 $& -459.9018425 & -79.880946 \\
&$6 $     & -459.7691030 & -79.758821 && &$5 $& -459.9024779 & -79.870426 \\
&$7 $     & -459.7692978 & -79.757749 && &$6 $& -459.9028828 & -79.869798 \\
&$8 $     & -459.7693594 & -79.757970 && &$7 $& -459.9031145 & -79.870329 \\
&$9 $     & -459.7694374 & -79.756594 && &$8 $& -459.9032110 & -79.871590 \\
&$12$     & -459.7695755 & -79.752764 && &$9 $& -459.9035007 & -79.863268 \\
&$15$     & -459.7695952 & -79.752384 && &$12$& -459.9038527 & -79.857397 \\
&         &              &            && &$14$& -459.9039916 & -79.855950 \\
&         &              &            && &$16$& -459.9040884 & -79.854589 \\
&         &              &            && &$20$& -459.9042568 & -79.850471 \\
&         &              &            && &$22$& -459.9042933 & -79.850431 \\
&         &              &            && &$27$& -459.9043424 & -79.849504 \\
&         &              &            && &$33$& -459.9043545 & -79.849094 \\
\br
\end{tabular}
\endTable
Considering that the results of the calculations performed on Cl and Cl$^-$ are independent would lead to separate studies of the convergence on each system and the summation of the so-deduced error estimations would give an overestimated final uncertainty on the theoretical $\EA$ and corresponding IS. This procedure indeed does not take account of the fact that the correlation contributions to the separate systems tend to cancel out when estimating the differential effects. To extract the differential $\Delta S_{\text{\sc{sms}}}$ value from our results, we need some guideline. In valence calculations, this guideline is the number of correlation layers. Unfortunately, in the open-core CI approach, there is no satisfactory intrinsic convergence path in the space of the calculation parameters $r$, $r'$, $p$ and $p'$.
Thanks to the correlation observed between the energy and the mass polarization expectation value~\cite{Caretal:2010a,CarGod:2011a}, we  use the reference $\EA_{\text{\scriptsize ref}}^\text{\scriptsize NR}$ estimated in Section~\ref{sec:EA_guideline} as a guideline.

\clearpage

\begin{figure}
\center
\caption{$E$ versus $S_{\text{\sc{sms}}}$ (in a.u.) plots of the results of the open-core CI calculations (see Table~\ref{tab5:fullCV}) for Cl$^-$ (a) and Cl (b). Convergence is read from right to left (decreasing energy). The plain line of Figure (a) is the relation~(\ref{eq:rel}) and the dashed lines correspond to the uncertainty on the fitted parameters (see text).\label{fig:CV}}
\hspace*{-3mm}\includegraphics[width=0.52\textwidth]{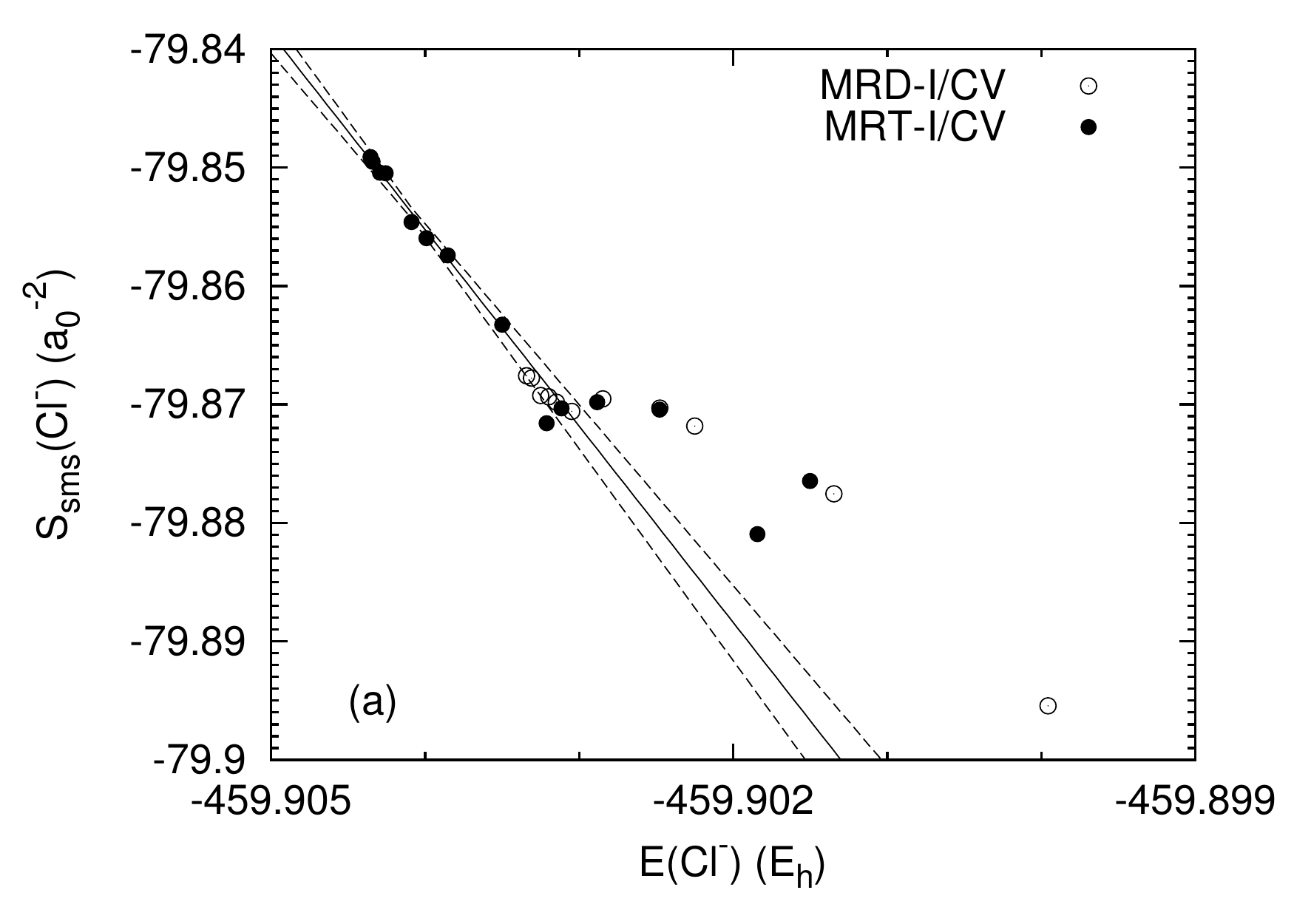}\hspace*{-5mm}
\includegraphics[width=0.52\textwidth]{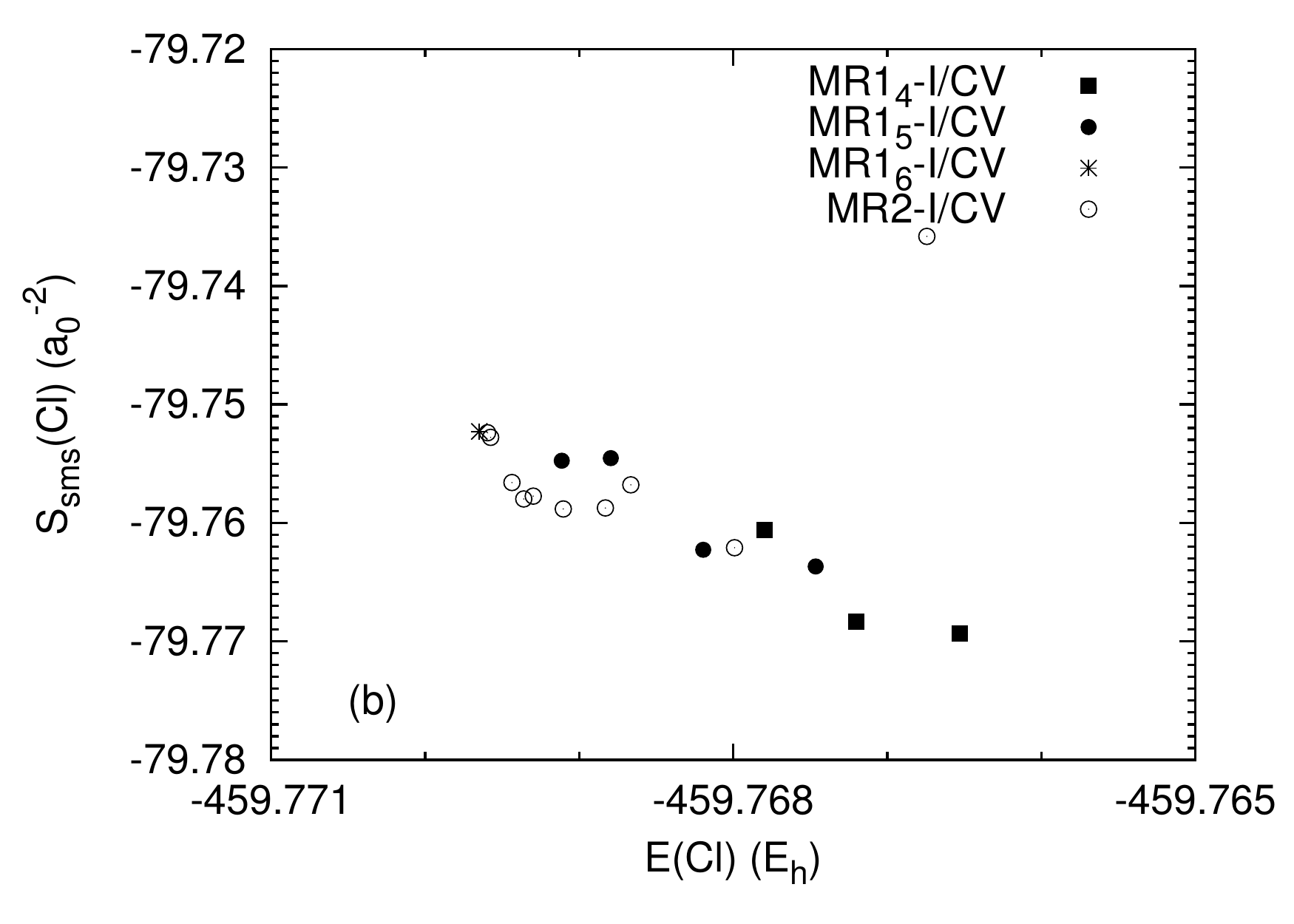}
\end{figure}

Figure~\ref{fig:CV} plots the $S_{\text{\sc{sms}}}$ results obtained in the $(r,p)$ model ($S_{\text{\sc{sms}}}(r,p)$), versus the energies $E(r,p)$, for Cl and Cl$^-$ in Figures (a) and (b) respectively. The $p$-values are given in Table~\ref{tab:ps}. Note that the same scale is used in both plots. The $p=1$ points are outside the frames. In order to extrapolate the results of the MR$1_5$ model, we add the configuration $3s^1 3p^4 4s^1 4p^1$ to MR1$_5$, generate the corresponding open-core MR1$_6$-I/CV$_6$$\lceil 11k \rceil$ CSF list and perform a CI calculation on this space using the MR1$_5$-I orbital set.

Figure~\ref{fig:CV}-a exhibits a good agreement of the results using the two references of Cl$^-$. Furthermore, the
convergence in $p$ (from right to left in the plots) is remarkably clean. Even though the correlation between $E(r,p)$ and $S_{\text{\sc{sms}}}(r,p)$ of Cl calculations is apparent in Figure~\ref{fig:CV}-b, it is less precise.

In these two plots, we see that $(\hbar^2/m_e) S_{\text{\sc{sms}}}$ is much more sensitive to $p$ than the total energy. Furthermore, we see from the comparison between Tables~\ref{tab:val1} and~\ref{tab:val2}, and Table~\ref{tab5:fullCV} that the $S_{\text{\sc{sms}}}$ parameter is much more affected than the total energy $E$ when open-core configurations are included in the expansion. This observation is supported by the larger impact of the core relaxation on the $S_{\text{\sc{sms}}}$ than on the energy in valence MR-I models (see Table~\ref{tab:val1}).
Fitting the eight last points based on the MRT model of Cl$^-$ (see Tables~\ref{tab5:fullCV}), one obtains (in atomic units)
\beq
S_{\text{\sc{sms}}}(\text{Cl}^-) = -16.6(7)\ \left[E(\text{Cl}^-) - \bar E \right] -79.8538(2)\label{eq:rel}
\eeq
where $\bar E=-459.904~085$~E$_h$ is the average of the energies of the fitted points.
If all the aligned points of Figure~\ref{fig:CV}-a are fitted, \emph{i.e.} if the five last points for MRD are also included in the fit, the angular coefficient becomes $-17.5(4)$. It gives an indication of the uncertainty of the fit considering the convergence with respect to the used MR ($r'$). The relation~(\ref{eq:rel}) is illustrated by a plain line in Figure~\ref{fig:CV}-a, the dashed lines being the lower and upper limits for an $1.3$ uncertainty on the angular coefficient and a $2.10^{-4}~a_0^{-2}$ uncertainty on the offset.
Using equation~(\ref{eq:rel}) with these upgraded uncertainties allows to extrapolate the Cl$^-$ energy and corresponding $S_{\text{\sc{sms}}}$ 
\beqa
\Delta S_{\text{\sc{sms}}} = S_{\text{\sc{sms}}}(\text{Cl}) + 16.6(13)\ \left[E(\text{Cl}) - \bar E - \EA_{\text{\scriptsize ref}}^\text{\scriptsize NR}\right] +79.8538(2) \nonumber\\\label{eq:extrap}
\eeqa
taking  
$\EA_{\text{\scriptsize ref}}^\text{\scriptsize NR} = 0.1347(1)$~E$_h$
from eq.~\eref{eq:EA_ref}. The convergence of the total weight of the $p$ first configurations of MRT-based wave functions with the corresponding energies suggests that, down to an energy of $-459.9052~$E$_h$, the variational principle is not violated within the limits of our model. 

Table~\ref{tab:extrap} gives the extrapolated energies and $S_{\text{\sc{sms}}}$ of the Cl$^-$ and the resulting $\Delta S_{\text{\sc{sms}}}$ values corresponding to each open-core CI calculation performed on Cl (see Table~\ref{tab5:fullCV}).
We regroup the results of the models with $r=1_4, 1_5$ and $1_6$ and order them by decreasing energy.
 The quoted errors only take into account the uncertainty deduced from the Cl$^-$ calculations convergence and from the estimated uncertainty on $\EA_{\text{\scriptsize ref}}^\text{\scriptsize NR}$.

In the $r=2$ calculations, the increase of the $S_{\text{\sc{sms}}}$(Cl) with $p$ is close to the angular coefficient of equation~(\ref{eq:rel}) giving $S_{\text{\sc{sms}}}$(Cl$^-$). It leads to the observed smoother convergence of the $r=2$ results of Table~\ref{tab:extrap} than the ones based on the other combined $r=1_4, 1_5, 1_6$ approaches. Still, all the calculations agree with the experiment ($\Delta S_{\text{\sc{sms}}}=0.091(25)~a_0^{-2}$). From Table~\ref{tab:extrap}, we adopt the value $0.096(3)~a_0^{-2}$. Using the fact that the results of $(r,p)=(2,15)$ and $(1_6,6)$ models are consistent, we estimate the uncertainty associated to the convergence of the calculations on Cl by taking the difference between the $(2,12)$ and $(1_5,5)$ values, leading to the final estimation $\Delta S_{\text{\sc{sms}}} = 0.096(9)~a_0^{-2}$.

\Table
{Total energies and $S_{\text{\sc{sms}}}$ parameters of Cl$^-$ and $\Delta S_{\text{\sc{sms}}}$ values extrapolated from the $(r,p)$ results of Cl to have \mbox{$E($Cl$^-)=E($Cl$) - {\EA}_{ref}^{NR}$}. $S_{\text{\sc{sms}}}($Cl$^-$) and $\Delta S_{\text{\sc{sms}}}$ are deduced from equation~(\ref{eq:extrap}). All values are in atomic units.\label{tab:extrap}}
\footnotesize
\begin{tabular}{crcccc}
\br
\multicolumn{2}{c}{Cl model} & \multicolumn{3}{c}{Cl extrapolation} \\
$r$ & $p$ & $E$(Cl$^-$)[E$_h$] & $S_{\text{\sc{sms}}}$(Cl$^-$)[$a_0^{-2}$] & $\Delta S_{\text{\sc{sms}}}$[$a_0^{-2}$]\\
\mr
1$_4$ & 1 & $-$459.8985 & $-$79.9456(98) & 0.1038(98)\\
1$_5$ & 1 & $-$459.8994 & $-$79.9312(86) & 0.0897(86)\\
1$_4$ & 2 & $-$459.9013 & $-$79.8995(59) & 0.1302(59)\\
      & 3 & $-$459.9020 & $-$79.8884(50) & 0.1201(50)\\
1$_5$ & 2 & $-$459.9022 & $-$79.8840(46) & 0.1203(46)\\
1$_4$ & 4 & $-$459.9026 & $-$79.8785(41) & 0.1179(41)\\
1$_5$ & 3 & $-$459.9030 & $-$79.8719(36) & 0.1096(36)\\
      & 4 & $-$459.9036 & $-$79.8619(27) & 0.1074(27)\\
      & 5 & $-$459.9039 & $-$79.8567(23) & 0.1019(23)\\
1$_6$ & 6 & $-$459.9044 & $-$79.8478(25) & 0.0955(25)\\      
\mr
2     & 1 & $-$459.8986 & $-$79.9437(96) & 0.1237(96)\\
      & 2 & $-$459.9015 & $-$79.8960(56) & 0.1602(56)\\
      & 3 & $-$459.9028 & $-$79.8753(39) & 0.1132(39)\\
      & 4 & $-$459.9034 & $-$79.8641(29) & 0.1073(29)\\
      & 5 & $-$459.9036 & $-$79.8614(27) & 0.1026(27)\\
      & 6 & $-$459.9039 & $-$79.8568(23) & 0.0980(23)\\
      & 7 & $-$459.9041 & $-$79.8536(21) & 0.0958(21)\\
      & 8 & $-$459.9041 & $-$79.8526(22) & 0.0946(22)\\
      & 9 & $-$459.9042 & $-$79.8513(23) & 0.0947(23)\\
      &12 & $-$459.9043 & $-$79.8490(24) & 0.0962(24)\\
      &15 & $-$459.9044 & $-$79.8487(25) & 0.0963(25)\\
\br
\end{tabular}
\endTable

\clearpage

\section{Comparison with experiment and conclusion}\label{sec:comp}

In Section~\ref{sec:valcsf}, we elaborate large scale valence MCHF models and subsequently add open-core CSFs expansions. Focusing on differential effects, more flexibility is added in the Cl$^-$ models than in the ones of Cl.
The correlation between the energy and the mass polarization expectation value convergences in a sequence of calculations, already observed in earlier work~\cite{CarGod:2011a}, allows to do  realistic extrapolations around the results of the largest Cl$^-$ calculations~(see Section~\ref{sec:cvres}). We hence fit our results to the experimental non-relativistic electron affinity estimated in Section~\ref{sec:EA_guideline}. A robust uncertainty for the (valence$ + $core-valence) correlation contributions to the isotope shift is deduced. However, core-core correlation effects are known to be important in isotope shifts calculations~\cite{MarSal:82a,Jonetal:99a} and more accurate approaches should be used, ultimately including relativistic corrections.

Table~\ref{tab:comp} compares results of our work with the experimental and theoretical values of Berzinsh \etal \cite{Beretal:95a}\footnote{Note that in their equation~(10) expressing the separation between the mass shift and the field shift, Berzinsh \etal adopted the first term of \eref{eq:ea_sep} for the mass contribution.}.
 The experimental normal mass shift (NMS) of $741$~MHz is used to deduce the total IS so that a cleaner theory/experiment comparison is found in the residual isotope shift ($\textrm{RIS}=\textrm{IS}-\textrm{NMS}$). The uncorrelated Dirac-Fock (DF) results from Berzinsh \etal\cite{Beretal:95a} are also reported. For each model, we give the specific mass shift (SMS), total mass shift ($\textrm{MS}=\textrm{NMS}+\textrm{SMS}$) and field shift (FS). 
 As pointed out by Berzinsh \etal\cite{Beretal:95a}, their estimated field isotope shift of the electron affinity is well below the error bars of the IS measurement. The difference between their FS values  and ours is caused by another choice of $\delta \la r^2\ra$ change between isotopes (see discussion in Section~\ref{sec5:FS}). With this respect, the uncertainty associated with the very small FS constitutes a large source of uncertainty in the final RIS and IS values.

 The calculations of Ref.~\cite{Beretal:95a} include the relativistic effects but only the lowest order correlation corrections, predicting a RIS opposite to the measured one. It is clearly seen from the \textsc{Fig}.~6 of this article displaying the behaviour of the SMS for different levels of approximation, that the theory-observation discrepancy is due to an insufficient treatment of the correlation effects.

\fulltable{Experimental and theoretical isotope shifts (IS) on the $\EA$ for the chlorine isotopes $A=37,35$. The experimental NMS of $0.741$~GHz is used to obtain the IS. All values are in GHz. Valence MCHF and FC-MCHF are based on the models MRD and MR2 respectively for Cl$^-$ and Cl. The final value is deduced as explained in Section~\ref{sec:cvres}.
  \label{tab:comp}}
  \footnotesize
\begin{tabular}{llllll}
 \br
        & \multicolumn{1}{c }{\hspace*{0.cm} SMS \hspace*{0.5cm} }
        & \multicolumn{1}{c }{\hspace*{0.cm} MS  \hspace*{0.5cm} }  
        & \multicolumn{1}{c }{\hspace*{0.cm} FS \hspace*{0.5cm} }  
        &  \multicolumn{1}{c }{\hspace*{0.cm} RIS\hspace*{0.5cm} } 
        &  \multicolumn{1}{c }{\hspace*{0.cm} IS \hspace*{0.5cm}} \\
\mr
&\multicolumn{5}{c}{This work}\\
\raisebox{3ex}{}%
HF & $-$1.348 & $-$0.607& $-$0.003(22) & $-$1.351(22) & $-$0.610(22)\\ 
val. FC$-$MCHF & $-$0.674 & +0.067 & $-$0.002(20) & $-$0.676(20) & +0.065(20)\\
val. MCHF & $-$0.495 & +0.246 & $-$0.003(21)  &  $-$0.497(21)& +0.244(21) \\
final results & $-$0.535(51) & +0.206(51) & $-$0.003(22)  & $-$0.538(72) & +0.203(72)\\
\raisebox{3ex}{}%
&\multicolumn{5}{c}{Berzinsh \etal \cite{Beretal:95a}}\\
\raisebox{3ex}{}%
Exp.   &   & &  & $-$0.51(14)& +0.22(14) \\
\raisebox{3ex}{}%
DF & $-$1.3 & $-$0.6 & +0.014(14) & $-$1.3& $-$0.6  \\
MB low corr.  & +0.50  & +1.24 & +0.014(14) & +0.51(2)& +1.26(2) \\
 \br
 \end{tabular}
\endfulltable

The results of the MRD-I$\lceil 11k\rceil$ for Cl$^-$ and MR2-I$\lceil 11k\rceil$ obtained in Section~\ref{sec:valres} are combined to give the valence frozen core (FC-MCHF) and optimized core (MCHF) estimations presented in Table~\ref{tab:comp}. At first sight, relaxing the core orbitals lifts the discrepancy between the valence model and the experimental values.
However, in this work, we observe that the optimization of the core orbitals of Cl and Cl$^-$ leads to significantly different contributions, \emph{a priori} destroying the balance between the core/valence distinction of the two systems. This effect arises from the occupation numbers symmetry in the Cl$^-$ main configuration. If it might be a good reason to distrust the closed-core MCHF values of Table~\ref{tab:comp}, it is not a demonstration that they should be ignored.

The isotope shift on the electron affinity is found to be anomalous for beryllium~\cite{Nemetal:04a} and oxygen~\cite{GodFro:99b,Bloetal:01a},  corresponding to a smaller electron affinity for the heavier isotope. For sulfur, the theoretical estimations also predict an anomalous~IS, allowed by the large experimental error bars, although the latter are  centered on a positive difference
( $\EA (^{34}\mbox{S}) -  \EA (^{32}\mbox{S}) = +0.0023(70)~\mbox{cm}^{-1}$)~\cite{Caretal:2010a}. The
IS on the  detachment thresholds were also found to be negative in carbon~\cite{CarGod:2011a}, waiting for some experimental confirmation. The isotope shift on the chlorine electron affinity is definitely normal, the NMS largely dominating the SMS. For all these systems, the FS are at least two orders of magnitude smaller than the total mass shift, to put in contrast with heavy negative ions, such as Os$^-$ for which 
the field shift clearly dominates the IS on the electric-dipole bound-bound transition frequency~\cite{Keletal:2011a}. For systems for which the FS can be neglected, the isotope shift on the electron affinity can be rationalized by looking at equation~\eref{eq:ea_sep}. The electron affinity being positive for a bound negative ion, the isotope shift will be normal if $\EA > (\hbar^2/m_e) \Delta S_{\text{\sc{sms}}}$. Taking the atoms along the third period, from Al to Cl, as an example, the electron affinity increases regularly from the left to the right of the periodic table, with a dip corresponding to phosphorous, while the  $\Delta S_{\text{\sc{sms}}}$ positive values defined by eq.~\eref{Delta_SMS} are relatively stable. Aluminum and silicon having a rather small $\EA$, they are expected to have a significant negative (anomalous) isotope shift while chlorine, as demonstrated in the present work, is characterized by a large electron affinity that determines the (normal) IS. The normal-anomalous IS change corresponds to the crossing between the $(\hbar^2/m_e) \Delta S_{\text{\sc{sms}}}$ and $\EA$ trends occuring around sulfur ($Z=16$) which IS sign on the electron affinity is not fully definite~\cite{Caretal:2010a}.

As pointed out by Andersson \etal~\cite{Andetal:2007a}, the analysis of the electron affinities of atoms shows that, with the exception of the hydrogen atom, calculated electron affinities are less accurately known than measured ones. However, we have  shown in~\cite{Caretal:2010a} that the theoretical IS values on the electron affinity could be more accurate than the observed ones. In the present work, we demonstrated that theoretical calculations of the electronic factors of the mass contribution  have evolved remarkably, solving a longstanding discrepancy between calculated and measured specific mass shift on the electron affinity of chlorine.

\ack

This work was supported by the Communaut\'e fran\c{c}aise of Belgium (Action de Recherche Concert\'ee), the Belgian National Fund for Scientific Research (FRFC/IISN Convention) and by the IUAP -- Belgian State Science Policy (BriX network P7/12).

\clearpage

\section*{References}
\bibliographystyle{unsrt}

\end{document}